\documentclass[prl,a4paper,12pt]{article}
\pdfoutput=1

\usepackage{bm}        						
\usepackage{amsmath}						
\usepackage{amssymb}						
\usepackage{amsthm}						    
\usepackage{array}							
\usepackage[english]{babel}					
\usepackage{booktabs}						
\usepackage{caption}						
\usepackage{cancel}							
\usepackage{color}                          
\usepackage[format=hang]{caption}			
\usepackage{eufrak}							
\usepackage[hang,flushmargin]{footmisc}		
\usepackage{fullpage}						
\usepackage{graphicx}						
\usepackage{hyperref}						
\hypersetup{
    colorlinks=true, 						
    linktoc=all,     						
    linkcolor=black,  						
    citecolor=black,
    filecolor=black,
    urlcolor=black
}

\usepackage{pdflscape}						
\usepackage{mathtools}						
\usepackage{multirow}						
\usepackage{nccmath}						
\usepackage{nicefrac}						
\usepackage[percent]{overpic}				
\usepackage[section]{placeins}				
\usepackage{relsize}						
\usepackage{setspace}						
\usepackage{slashed}						
\usepackage{subcaption}						
\usepackage{xcolor}                         

\numberwithin{equation}{section}				
\newcommand{\Naturals}{\mathbb{N}}			
\newcommand{\Reals}{\mathbb{R}}				
\newcommand{\Cyclic}{\mathbb{Z}}		        

\makeatletter
\newcommand*{\defeql}{\mathrel{\rlap{%
                     \raisebox{0.3ex}{$\m@th\cdot$}}%
                     \raisebox{-0.3ex}{$\m@th\cdot$}}%
                     =}
\makeatother								

\makeatletter
\newcommand*{\defeqr}{=\mathrel{\rlap{%
                     \raisebox{0.3ex}{$\m@th\cdot$}}%
                     \raisebox{-0.3ex}{$\m@th\cdot$}}%
                     }
\makeatother								

\begin{document}

\thispagestyle{empty}

\vspace*{2cm}

\begin{center}
   {\LARGE \bf Perturbative Construction of Stationary\\Randall-Sundrum II Black Holes\\on a 5-Brane\par}
    \vspace{2.5cm}

{\bf Maren Stein$^\dagger$}

\vspace*{1cm}

 {\it $\dagger$ DAMTP, Centre for Mathematical Sciences, University of Cambridge,\\
 Wilberforce Road, Cambridge CB3 0WA, United Kingdom}

\vspace*{0.5cm} {\tt mcs60@cam.ac.uk}
\vspace{0.5cm} {date}
\end{center}

\vspace*{1cm}

\begin{abstract}
We numerically construct large Randall-Sundrum II brane black holes in 4 and 5 dimensions from associated AdS/CFT spacetimes. Our solutions are leading order perturbations of a representative of the boundary conformal structure of the AdS spacetime sourced by the dual CFT stress tensor. The 4-dimensional solutions are static perturbations of the Euclidean Schwarzschild metric, while the 5-dimensional solutions are perturbations of the Myers-Perry metric with equal angular momenta. We compare the former with previous numerical results for Randall-Sundrum bulk black holes and find good agreement down to a horizon radius of about $r_H\sim30\ell$. The latter are the first numerical results pertaining to rotating Randall-Sundrum black holes. They have the same entropy, but a larger horizon area than Myers-Perry black holes of the same mass and angular momentum. 
\end{abstract}
\noindent

\newpage
\setcounter{tocdepth}{2} 
\tableofcontents

\setcounter{page}{1} \setcounter{footnote}{0}

\section{Introduction}
The Randall-Sundrum II (RS2) model~\cite{Randall:1999vf} consists of a single brane embedded in a mirror-symmetric AdS bulk. In the low energy limit it yields effectively 4-dimensional physics on the brane despite the infinite extra dimension. Using AdS/CFT arguments it has been conjectured that the low energy theory on the brane is gravity coupled to a conformal field theory (CFT)~\cite{Gubser:1999vj, Duff:2000mt,Verlinde:1999fy,Verlinde:1999xm,Giddings:2000mu,Emparan:2002px}. To be phenomenologically relevant the RS2 model must admit realistic black hole solutions. So far, analytic approaches have not produced a line element describing a stable, regular 4-dimensional RS2 brane world black hole. Though, significant progress has been made numerically. It had been claimed that large RS2 black holes, with horizon radius larger than the AdS length, would quickly evaporate as the degrees of freedom of the coupled CFT provide a large number of additional channels for Hawking radiation~\cite{Tanaka:2002rb,Emparan:2002px}. This argument, however, is based on free-field intuition whereas the CFT is strongly coupled~\cite{deHaro:2000wj}. The dispute was settled when Figueras and Wiseman presented static, stable numerical black hole solutions in 5 dimensions with horizon radii up to $R_H=100\ell$~\cite{Figueras:2011gd,Figueras:2014nza}. Recently,~\cite{Wang:2015rka,Wang:2016nqi} performed the first numerical simulation of gravitational collapse in the RS2 scenario. For sufficiently strong initial conditions the computations yield black holes. These spherical solutions agree well with the results of Figueras and Wiseman. Evidence suggests that the black holes resulting from gravitational collapse do not depend on details of the in initial data. Gravitational equations on the brane which involve a dual CFT stress energy tensor and higher oder curvature corrections have been derived in~\cite{deHaro:2000wj}. Based on these~\cite{Figueras:2011gd} showed how a low energy RS2 solution can be constructed as perturbation of a representative of the boundary conformal structure of an associated asymptotically AdS spacetime. We applied this perturbative construction to leading order to the static $AdS_5/CFT_4$ solution of~\cite{Figueras:2011va} and the stationary $AdS_6/CFT_5$ solution of~\cite{Figueras:2013jja}. This perturbative approach has been employed previously in~\cite{Abdolrahimi:2012qi,Abdolrahimi:2012pb}. Using a numerical method independent of the Ricci-flow approach of~\cite{Figueras:2011va} the authors constructed an $AdS_5/CFT_4$ solution that is asymptotically conformal to the Schwarzschild metric and perturbed it to leading order. The resulting brane black holes were found to agree well with the full solutions of~\cite{Figueras:2011gd}. Our calculations yield the first numerical results for rotating RS2 brane black holes and provide an estimate of the validity range of the leading order perturbative approach.\\
This paper is structured as follows: In section~\ref{sec: perturbative construction} we briefly review the perturbative construction of low curvature RS2 spacetimes from associated AdS/CFT solutions. Section~\ref{sec: ansatz} details the ansatz and analytic results of our calculations, while the numerical results are presented in section~\ref{sec: numerical results}. We use Planck units $c=\hbar=k=G=1$, except in sections~\ref{sec: perturbative construction}, and~\ref{subsec: metric ansatz} and in appendix~\ref{app: grav eqs brane}, where the gravitational constants have been left general.
\section{Perturbative Construction of Low Energy Randall-Sundrum II Spacetimes from AdS/CFT Solutions}\label{sec: perturbative construction}
This section follows~\cite{Figueras:2011gd}, though the number of dimensions has been kept general. The derivation uses the gravitational equations on the brane derived in~\cite{deHaro:2000wj}, which are briefly reviewed in appendix~\ref{app: grav eqs brane}.\\
The principal idea is to slice a perturbation of a known AdS/CFT solution with a brane close to the conformal boundary. The perturbation can be chosen such that it accounts for the gravitational back-reaction of the brane. Then two copies of the truncated spacetime joined along their common boundary constitute a solution to the gravitational equation on the brane~\eqref{RS2 brane grav eq_AdS/CFT}. We will only consider flat RS2 branes, where the brane tension is fine tuned such that the cosmological constant on the brane~\eqref{cosmo const brane AdS/CFT} vanishes. It is convenient to rewrite equation~\eqref{RS2 brane grav eq_AdS/CFT} in terms of the rescaled brane metric $\tilde{g}_{\mu\nu}=\frac{\epsilon^2}{\ell^2}\gamma_{\mu\nu}$:
\begin{equation}\label{grav eq brane gtilde}
G_{\mu\nu}[\tilde{g}]+\mathcal{O}\left(R[\tilde{g}]^2\right)=\left(\frac{\epsilon}{\ell}\right)^{d-2}8\pi G_{d}\left[2\langle T^{\text{CFT}}_{\mu\nu}[\tilde{g}]\rangle+\tau_{\mu\nu}\right]\,,
\end{equation}
Now consider a $(d+1)$-dimensional asymptotically AdS spacetime $\mathfrak{M}$ that, far from the conformal boundary, tends to the Poincar\'{e}e horizon of AdS. We are interested in the near boundary region where the Fefferman-Graham expansion~\eqref{FG expansion} is valid. The expansion on $\mathfrak{M}$ is completely determined by a representative $\mathfrak{g}^{(0)}_{\mu\nu}$ of the boundary conformal structure and a symmetric tensor $\mathfrak{t}_{\mu\nu}$, which is related to the dual CFT stress-energy tensor via~\eqref{general form of stress tensor}. Assume that, for sufficiently small $\epsilon$, there exist asymptotically AdS spacetimes $M_\epsilon$ whose boundary metric is a perturbation of the boundary metric of the original spacetime $\mathfrak{M}$, i.e., $g^{(0)}_{\mu\nu}(\epsilon)=\mathfrak{g}^{(0)}_{\mu\nu}+H_{\mu\nu}(\epsilon)$. For two copies of $M_\epsilon$, sliced by a brane at $z=\epsilon$, joined together along their common boundary and identified via a $\Cyclic_2$-symmetry, to produce an RS2 solution the rescaled brane metric $\tilde{g}_{\mu\nu}(\epsilon)$ has to obey equation~\eqref{grav eq brane gtilde}. Inserting $g^{(0)}_{\mu\nu}(\epsilon)$ into the near boundary expansion~\eqref{FG expansion} shows that $\tilde{g}_{\mu\nu}(\epsilon)$ itself can be written as a perturbation of the form $\tilde{g}_{\mu\nu}(\epsilon)=\mathfrak{g}^{(0)}_{\mu\nu}+h_{\mu\nu}(\epsilon)$. Solving equation~\eqref{grav eq brane gtilde} order by order in $\epsilon$ will determine the series expansion  
\begin{equation}
h_{\mu\nu}(\epsilon)=h^{(0)}_{\mu\nu}+\left(\frac{\epsilon}{\ell}\right)^2h^{(2)}_{\mu\nu}+\ldots+\left(\frac{\epsilon}{\ell}\right)^{d-2}h^{(d-2)}_{\mu\nu}+\ldots\,,
\end{equation}
where only even powers of $\epsilon$ appear up to order $(d-2)$ and in even dimension the series contains a logarithmic term. If the brane carries no matter, i.e., $\tau_{\mu\nu}=0$, $h^{(0)}_{\mu\nu}=0$ and $\tilde{g}_{\mu\nu}(\epsilon)$ is Ricci flat to oder $\epsilon^0$. Then all curvature terms up to order $\epsilon^{d-2}$ vanish and the leading correction stems from the dual CFT stress-energy tensor of $M_\epsilon$. The additional assumption $t_{\mu\nu}(\epsilon)=\mathfrak{t}_{\mu\nu}+\mathcal{O}(\epsilon^{d-1})$ allows one to replace $\langle T^{\text{CFT}}_{\mu\nu}[\tilde{g}]\rangle$ by $\langle T^{\text{CFT}}_{\mu\nu}[\mathfrak{g}^{(0)}]\rangle$. Since now $G_{\mu\nu}[\tilde{g}]=G_{\mu\nu}[\mathfrak{g}^{(0)}]+\left(\frac{\epsilon}{\ell}\right)^{d-2}G_{\mu\nu}^{(d-2)}[\mathfrak{g}^{(0)},h^{(d-2)}]+\mathcal{O}\left(\epsilon^{d-1}\right)$ equation~\eqref{grav eq brane gtilde} yields at leading order $\epsilon^{d-2}$
\begin{equation}\label{central eq}
\boxed{
G_{\mu\nu}^{(d-2)}[\mathfrak{g}^{(0)},h^{(d-2)}]=16\pi G_d\,\langle T^{\text{CFT}}_{\mu\nu}[\mathfrak{g}^{(0)}]\rangle
}\,.
\end{equation}
Our aim is to solve equation~\eqref{central eq}, which is merely the linearized Einstein's equation, for two cases, where $\mathfrak{M}$ is identified with the AdS/CFT solutions of~\cite{Figueras:2011va} and~\cite{Figueras:2013jja}. Their boundary metrics are conformal to the 4-dimensional Euclidean Schwarzschild metric and the 5-dimensional Myers-Perry metric with equal angular momenta. Our calculations will yield the brane metric $\gamma_{\mu\nu}=\frac{\ell^2}{\epsilon^2}\tilde{g}_{\mu\nu}(\epsilon)$ to leading order in $\epsilon$, but not the full bulk solution. As the boundary metric $\mathfrak{g}_{\mu\nu}^{(0)}$ of the asymptotically AdS spacetime is the background for the perturbation $h_{\mu\nu}^{(d-2)}$ we will also refer to it as background metric.
\section{Ansatz and Analytic Results}\label{sec: ansatz}
\subsection{Brane Metric}\label{subsec: metric ansatz}
\subsubsection{Static 4-dimensional Brane Black Holes}
The Euclidean Schwarzschild metric is a representative of the conformal class of the $AdS_5/CFT_4$ spacetime~\cite{Figueras:2011va}. Its $U(1)\times SO(3)$ isometry is apparent from the line element 
\begin{equation}\label{4D bd metric}
ds^2_{\text{ESS}}=\left(1-\frac{R_0}{R}\right)d\tau^2+\left(1-\frac{R_0}{R}\right)^{-1}dR^2+R^2d\Omega_{(2)}^2\,,
\end{equation}
where $R_0=2M$ is the horizon radius of the black hole horizon with surface gravity $\kappa=\frac{1}{2R_0}$. After the introduction of a compact radial coordinate via
\begin{equation}\label{4D compact coordinate}
R=R_0/\left(1-r^2\right)
\end{equation}
and a convenient rescaling of the time coordinate with a factor $R_0$ the line element reads
\begin{equation}
ds^2_{\text{ESS}}=r^2R_0^2d\tau^2+\frac{4R_0^2}{\left(1-r^2\right)^4}dr^2+\frac{R_0^2}{\left(1-r^2\right)^2}d\Omega_{(2)}^2\,.
\end{equation}
The horizon now lies at $r=0$ while one approaches spatial infinity as $r\rightarrow 1$. The dual stress-energy tensor has the form\footnote{Note that in equation (4.18) of~\cite{Figueras:2011va} the pre-factor of the stress tensor involves the number of colours $N_c$ of the dual CFT, which is related to the 5-dimensional Newton constant via $\frac{\ell^3}{8\pi G_5}=\left(\frac{N_c}{2\pi}\right)^2$.}
\begin{equation}\label{4D CFT stress tensor}
\begin{fleqn}
\begin{aligned}
\left\langle{T^{\rm CFT}}_\mu^\nu\right\rangle=\frac{\ell^3}{4\pi G_5}\frac{1}{R^4}\text{diag}\biggl\{&\frac{3R_0}{4R}\left(1-\frac{R_0}{R}\right)+t_4(R),\frac{3R_0^2}{4R^2}-\left(2s_4(R)+t_4(R)\right),\\
&-\frac{3R_0}{8R}+s_4(R),-\frac{3R_0}{8R}+s_4(R)\biggr\}\,.
\end{aligned}
\end{fleqn}
\end{equation}
The stress tensor is traceless because the coefficient $a^{(d)}$ in the expansion~\eqref{FG expansion} vanishes for a Ricci-flat boundary metric. As the stress tensor is covariantly conserved the functions $s_4$ and $t_4$ are not independent, but obey a constraint. The CFT stress tensor shares the isometries of the boundary metric and in turn imposes them on the brane metric. So we make the ansatz
\begin{equation}\label{ansatz 4D}
ds^2_4\tilde{g}_{\mu\nu}dx^\mu dx^\nu=r^2R_0^2T_4(r)d\tau^2+\frac{4R_0^2}{\left(1-r^2\right)^4}A_4(r)dr^2+\frac{R_0^2}{\left(1-r^2\right)^2}S_4(r)d\Omega_{(2)}^2\,,
\end{equation}
where
\begin{equation}
X_4(r)=X_4^{(0)}(r) \biggl(1+\left(\frac{\epsilon}{\ell}\right)^2\,X_4^{(2)}(r)\biggr),\quad X=T,\,A,\,S\,.
\end{equation}
With $T^{(0)}_4=A^{(0)}_4=S^{(0)}_4=1$ the brane metric approaches~\eqref{4D bd metric} as $\epsilon\rightarrow 0$.
\subsubsection{Stationary 5-dimensional Brane Black Holes}
The boundary conformal class of the $AdS_6/CFT_5$ spacetime~\cite{Figueras:2013jja} contains the Myers-Perry metric with equal angular momenta. Its $\Reals_t\times SU(2)\times U(1)$ isometry group is manifest in Boyer-Lindquist coordinates
\begin{equation}\label{5D MP non-compact}
ds^2_{\text{MP}_5}=-dt^2+\frac{R^2\left(R^2+a^2\right)}{\left(R^2+a^2\right)^2-\mu R^2}dR^2+\frac{\mu}{R^2+a^2}\left(dt+\frac{a}{2}\sigma^3\right)^2+\left(R^2+a^2\right)d\Omega_{(3)}^2\,.
\end{equation}
As usual, $\mu$ and $a$ denote the mass and angular momentum parameters. The line element above is written in a static frame, where the asymptotic flatness is manifest. The standard metric on the unit 3-sphere is $d\Omega_{(3)}^2=\frac{1}{4}\left(\left(\sigma^1\right)^2+\left(\sigma^2\right)^2+\left(\sigma^3\right)^2\right)$. The left-invariant two-forms of $SU(2)$ are given by
\begin{equation}\label{SU(2) forms sigma}
\begin{aligned}
\sigma^1=&-\sin\psi\,d\theta+\sin\theta\cos\psi\,d\phi\,,\\
\sigma^2=&\cos\psi\,d\theta+\sin\theta\sin\psi\,d\phi\,,\\
\sigma^3=&\,d\psi+\cos\theta\,d\phi\,,
\end{aligned}
\end{equation}
with $0\leq \theta\leq\pi,\;0\leq\phi\leq 2\pi$, and $0\leq\psi\leq 4\pi$. The largest real root of $\left(R^2+a^2\right)^2-\mu R^2=0$ determines the event horizon $R_H$. This can be used to express the mass parameter in terms of $a$ and $R_H$
\begin{equation}
\mu=\frac{\left(R_H^2+a^2\right)^2}{R_H^2}\,.
\end{equation}
The black hole's angular velocity and surface gravity are given by
\begin{equation}
\Omega_H=-\frac{2a}{R_H^2+a^2}\,,\hspace{50pt}\kappa^2=\frac{\left(R_H^2-a^2\right)^2}{R_H^2\left(R_H^2+a^2\right)^2}\,.
\end{equation}
After the introduction of a compact radial coordinate defined via
\begin{equation}\label{5D compact coordinate}
R^2+a^2=\frac{R_H^2+a^2}{\left(1-r^2\right)^2}\,
\end{equation}
the line element becomes
\begin{equation}
ds^2_{\text{MP}_5}=-r^2T_5^{(0)}(r)dt^2+\frac{4 A_5^{(0)}(r)}{\left(1-r^2\right)^4}dr^2+\frac{B_5^{(0)}(r)}{4\left(1-r^2\right)^2}\left(\sigma^3-\Omega_5^{(0)}(r)dt \right)^2+\frac{S_5^{(0)}(r)}{4\left(1-r^2\right)^2}d\Omega_{(2)}^2\,,
\end{equation}
with $d\Omega_{(2)}^2=\left(\sigma^1\right)^2+\left(\sigma^2\right)^2$ and
\begin{subequations}
\begin{align}
T^{(0)}_5(r)=&\frac{\left(2-r^2\right)\left(R_H^2-a^2\left(1-r^2\right)^2\right)}{R_H^2+a^2\left(1-r^2\right)^4}\,,\\
A^{(0)}_5(r)=&\frac{R_H^2\left(a^2+R_H^2\right)}{\left(2-r^2\right)\left(R_H^2-a^2\left(1-r^2\right)^2\right)}\,,\\
B^{(0)}_5(r)=&\,\frac{\left(a^2+R_H^2\right)\left(R_H^2+a^2\left(1-r^2\right)^4\right)}{R_H^2}\,,\\
\Omega^{(0)}_5(r)=&-\frac{2a\left(1-r^2\right)^4}{R_H^2+a^2\left(1-r^2\right)^4}\,,\\
S^{(0)}_5(r)=&\,a^2+R_H^2\,.
\end{align}
\end{subequations}
The dual stress-energy tensor is of the form\footnote{Note that there is a mistake in the expression for the stress-energy tensor in equation (3.8) of~\cite{Figueras:2013jja}.}
\begin{equation}\label{5D CFT stress tensor}
\begin{split}
\left\langle T^{\rm CFT}_{\mu\nu}\right\rangle dx^\mu dx^\nu=\frac{5\ell^4}{16\pi G_6}\biggl[&-T_{\rm CFT}(r)dt^2+A_{\rm CFT}(r)dr^2+B_{\rm CFT}(r)\left(\sigma^3-\Omega_5^{(0)}(r)dt \right)^2\\
&-\frac{B_5^{(0)}(r)}{2\left(1-r^2\right)^2}\left(\sigma^3-\Omega_5^{(0)}(r)dt \right)\Omega_{\rm CFT}(r)dt+S_{\rm CFT}\,d\Omega_{(2)}^2\biggr].
\end{split}
\end{equation}
The functions $X_{\rm CFT}$, $X=T,A,B,\Omega,S$ change with the angular momentum parameter $a$ of the boundary black hole, but are merely rescaled when the mass parameter $\mu$ varies. The stress tensor is covariantly conserved as well as traceless due to the absence of a conformal anomaly in odd dimensions. Consequently the functions $X_{\rm CFT},$ obey two constraints. As before, the brane metric shares the isometries of the boundary metric. So we make the ansatz
\begin{equation}\label{ansatz 5D}
\begin{aligned}
ds^2_5=\tilde{g}_{\mu\nu}dx^\mu dx^\nu=&-r^2T_5(r)dt^2+\frac{4 A_5(r)}{\left(1-r^2\right)^4}dr^2+\frac{B_5(r)}{4\left(1-r^2\right)^2}\left(\sigma^3-\Omega_5(r)dt \right)^2\\
&+\frac{S_5(r)}{4\left(1-r^2\right)^2}d\Omega_{(2)}^2\,,
\end{aligned}
\end{equation}
with
\begin{subequations}\label{5D form of perturbations}
\begin{align}
X_5(r)=&X_5^{(0)}(r) \biggl(1+\left(\frac{\epsilon}{\ell}\right)^3\,X_5^{(3)}(r)\biggr)\,,\qquad X=T,\,A,\,B,\,S,\\
\Omega_5(r)=&\Omega^{(0)}_5(r) \biggl(1+\left(\frac{\epsilon}{\ell}\right)^3\, \left(1-r\right)^{-1}\Omega_5^{(3)}(r)\biggr)\,.\label{5D perturbations Omega}
\end{align}
\end{subequations}
\subsection{Boundary Conditions and Choice of Gauge}\label{subsec: gauge & bcs}
Our task is to numerically solve the stationary linearized Einstein equations with an effective stress-energy tensor in 4 and 5 dimensions. For a certain class of black hole spacetimes, including the asymptotically flat case in higher dimensions, the stationary Einstein equations can be phrased as elliptic boundary value problem~\cite{Adam:2011dn}. Instead of the weakly elliptic Ricci tensor one considers the strongly elliptic operator
\begin{equation}
R^\text{H}_{\mu\nu}=R_{\mu\nu}-\nabla_{(\mu}\xi_{\nu)}\,,
\end{equation}
where $\xi^\mu$ is the DeTurck vector field~\cite{DeTurck:1983doe} defined with respect to a smooth reference metric $\bar{g}_{\mu\nu}$ as
\begin{equation}\label{DeTurck vec}
\xi_\mu=g^{\rho\sigma}\left(\bar{\nabla}_{\rho}g_{\sigma\mu}-\frac{1}{2}\bar{\nabla}_\mu g_{\rho\sigma}\right)\,,\quad\xi^\mu=g^{\rho\sigma}\left(\Gamma^\mu_{\rho\sigma}-\bar{\Gamma}^\mu_{\rho\sigma}\right)\,.
\end{equation}
As the difference of two connections the vector field is globally well defined. In this DeTurck gauge the gauge condition $\xi^\mu=0$ is not satisfied a priori, but solved simultaneously with the Einstein equations. Hence in the 4-dimensional case equation~\eqref{central eq} boils down to a system of 3 linear coupled second order ODEs for the functions $X_4^{(2)}$, while in the 5-dimensional case one is left with a system of 5 coupled ODEs for the functions $X_5^{(3)}$.\\
Boundary conditions are derived by demanding the metric be asymptotically flat and regular at the horizon and the axes of symmetry. For the metric~\ref{ansatz 4D} asymptotic flatness boils down two
\begin{equation}\label{4D bc infty}
X_4^{(2)}(r)|_{r=1}=0,\,\qquad X=T,\,A,\,S.
\end{equation}
While regularity at the horizon imposes Neumann conditions
\begin{equation}\label{4D bc horizon}
{X_4^{(2)}}'(r)|_{r=0}=0,\qquad X=T,A,S\,,
\end{equation}
as well as the additional condition
\begin{equation}\label{4D bc surfgrav}
A_4^{(2)}(r)|_{r=0}=T_4^{(2)}(r)|_{r=0}\,.
\end{equation}
For the metric~\ref{ansatz 5D} asymptotic flatness requires
\begin{equation}
X_5^{(3)}(r)|_{r=1}=0,\,\qquad X=T,\,A,\,B,\,\Omega,\,S.
\end{equation}
The factor of $(1-r)^{-1}$ multiplying $\Omega^{(3)}_5$ in equation~\eqref{5D form of perturbations} is crucial to allow a contribution of order $R^{-2}$ to $g_{t\psi}$ in the asymptotic region and thereby a change in the angular momentum. Regularity at the horizon requires
\begin{subequations}
\begin{align}
{X_5^{(3)}}'(r)|_{r=0}=0,&\qquad X=T,\,A,\,B,\,\Omega,\,S,\\
A_5^{(3)}(r)|_{r=0}=T_5^{(3)}(r)|_{r=0}\,,&\qquad\Omega_5^{(3)}(r)|_{r=0}=0.\label{5D bc surfgrav}
\end{align}	
\end{subequations}
Note that conditions~\eqref{4D bc surfgrav} and~\eqref{5D bc surfgrav} ensure that the surface gravity and the angular velocity of the horizon remain unchanged. We performed our analytic calculations in both, the standard transverse traceless gauge and the DeTurck gauge. The former, however, turned out to not be well-behaved for the specific ansatz we chose for the brane metric.  
\subsection{Physical Quantities}
Like the brane metric,
\begin{equation}
\gamma_{\mu\nu}=\left(\frac{\ell}{\epsilon}\right)^2\Bigl[g_{\mu\nu}^{(0)}+\left(\frac{\epsilon}{\ell}\right)^{d-2}h_{\mu\nu}^{(d-2)}+\mathcal{O}\left(\epsilon^{d-1}\right)\Bigr]\,,
\end{equation}
itself a quantity $Q$ computed from it will be of the form
\begin{equation}
Q[\gamma]=\left(\frac{\ell}{\epsilon}\right)^{k}\Bigl[Q^{(0)}[g^{(0)}]+\left(\frac{\epsilon}{\ell}\right)^{d-2}Q^{(d-2)}[g^{(0)},h^{(d-2)}]+\mathcal{O}\left(\epsilon^{d-1}\right)\Bigr]\,,
\end{equation}
where $k$ depends on the scaling dimension of $Q$. The scaling factor $\ell/\epsilon$ carries no physical meaning. It can be eliminated by rescaling quantities with the correct power of the brane black hole mass $M$ in order to make them dimensionless.\\ 
Gravity on the RS2 brane is effectively lower-dimensional and as the leading order corrections to the gravitational potential~\cite{Garriga:1999yh} fall-off faster than $\mathcal{O}\left(1/r^{d-2}\right)$ the black hole mass is given by the ADM mass. In both cases we consider the black hole mass receives no perturbative correction. So, to leading order, one finds
\begin{equation}\label{brane BH mass}
M=\left(\frac{\ell}{\epsilon}\right)^{d-3}M^{(0)}_d\,,\qquad d=4,5,
\end{equation}
with $M^{(0)}_4=R_0/2$ and $M^{(0)}_5=3\pi\left(a^2+R_H^2\right)^2/8R_H^2\,$. The static brane black holes are distinguished by their horizon radius, which, to leading order, reads
\begin{equation}\label{4D brane horizon radius}
R_4=\frac{\ell}{\epsilon}R_0\left(1+\frac{1}{2}\left(\frac{\epsilon}{\ell}\right)^2S_4^{(2)}(0)\right).
\end{equation}
The stationary solutions are parametrized by their horizon radius and angular momentum. To leading order the radius is given by
\begin{equation}\label{5D brane horizon radius}
R_5=\frac{\ell}{\epsilon} R_H\left(1+\frac{1}{2}\left(\frac{\epsilon}{\ell}\right)^3S_5^{(3)}(0)\right)\,,
\end{equation}
while the angular momentum can be determined from the Komar integral~\cite{Komar:1958wp,Jaramillo:2010ay} associated with the rotational Killing vector $k^\mu=\partial/\partial_\psi$ of the metric~\eqref{ansatz 5D}. The leading order result reads
\begin{equation}\label{5D brane angular momentum}
J_5=\left(\frac{\ell}{\epsilon}\right)^3J^{(0)}_5\left(1-\left(\frac{\epsilon}{\ell}\right)^3{\Omega_5^{(3)}}'(1)\right)\,,\qquad J^{(0)}_5=-\frac{a\pi\left(a^2+R_H^2\right)^2}{4R_H^2}\,.
\end{equation}
As gravity on the brane is not pure Einstein gravity, the Bekenstein-Hawking formula no longer holds. The first law of thermodynamics, however, remains valid. In the cases we consider only the angular momentum receives a perturbative correction.\footnote{Recall the result~\eqref{brane BH mass} and the fact that due to the boundary conditions~\eqref{4D bc surfgrav} and~\eqref{5D bc surfgrav} the surface gravity and the angular velocity of the horizon remain unchanged.} So the first law integrates to give the Smarr formula
\begin{equation}\label{Smarr formula}
\left(d-3\right)M=\left(d-2\right)\left(\frac{1}{8\pi}\kappa\mathcal{A}_H+\Omega_H J\right).
\end{equation}
The explicit results for the entropy are
\begin{alignat}{2}
\mathcal{S}_4=&\left(\frac{\ell}{\epsilon}\right)^2\mathcal{S}^{(0)}_4\,,\qquad &\mathcal{S}^{(0)}_4=&\pi R_0^2\,,\\
\mathcal{S}_5=&\left(\frac{\ell}{\epsilon}\right)^3\mathcal{S}^{(0)}_5\left(1+\left(\frac{\epsilon}{\ell}\right)^3\frac{2a^2}{R_H^2-a^2}\,{\Omega_5^{(3)}}'(1)\right)\,,\qquad &\mathcal{S}^{(0)}_5=&\frac{\pi^2\left(a^2+R_H^2\right)^2}{2R_H}\,.
\end{alignat} 
For the horizon area one finds
\begin{alignat}{2}
{\mathcal{A}_H}_4=&\left(\frac{\ell}{\epsilon}\right)^2\mathcal{A}_4^{(0)}\left(1+\left(\frac{\epsilon}{\ell}\right)^2S_4^{(2)}(0)\right)\,,\quad &\mathcal{A}_4^{(0)}=&4\pi R_0^2\,,\label{4D horizon area} \\
{\mathcal{A}_H}_5=&\left(\frac{\ell}{\epsilon}\right)^3\mathcal{A}^{(0)}_5\left(1+\frac{1}{2}\left(\frac{\epsilon}{\ell}\right)^3\left(B_5^{(3)}(0)+2S_5^{(3)}(0)\right)\right)\,,\quad &\mathcal{A}^{(0)}_5=&\frac{2\pi^2\left(a^2+R_H^2\right)^2}{R_H}\,.\label{5D horizon area}
\end{alignat}
In both cases the Ricci scalar $R^{(0)}$ and the Ricci tensor $R_{\mu\nu}^{(0)}$ of the background metric vanish. Hence, to leading order, the curvature scalars are of the form
\begingroup
\allowdisplaybreaks
\begin{subequations}\label{brane curvature scalars}
\begin{align}
R=&\left(\frac{\ell}{\epsilon}\right)^{-2}\left[R^{(0)}+\left(\frac{\epsilon}{\ell}\right)^{d-2}R^{(d-2)}\right]=\left(\frac{\epsilon}{\ell}\right)^dR^{(d-2)}\,,\\
R_{\mu\nu}R^{\mu\nu}=&\left(\frac{\ell}{\epsilon}\right)^{-4}\left[R^{(0)}_{\mu\nu}+\left(\frac{\epsilon}{\ell}\right)^{d-2}R^{(d-2)}_{\mu\nu}\right]\left[{R^{(0)}}^{\mu\nu}+\left(\frac{\epsilon}{\ell}\right)^{d-2}{R^{(d-2)}}^{\mu\nu}\right]\\ \nonumber
=&\left(\frac{\ell}{\epsilon}\right)^{-4}\left(\frac{\epsilon}{\ell}\right)^{2\left(d-2\right)}\left(R_{\mu\nu}R^{\mu\nu}\right)^{\left(2\left(d-2\right)\right)}=\left(\frac{\epsilon}{\ell}\right)^{2d}\left(R_{\mu\nu}R^{\mu\nu}\right)^{\left(2\left(d-2\right)\right)}\,,\\
K=&\left(\frac{\ell}{\epsilon}\right)^{-4}\left[R^{(0)}_{\mu\nu\rho\sigma}{R^{(0)}}^{\mu\nu\rho\sigma}+2\left(\frac{\epsilon}{\ell}\right)^{d-2}R^{(0)}_{\mu\nu\rho\sigma}{R^{(d-2)}}^{\mu\nu\rho\sigma}\right]\\ \nonumber
=&\left(\frac{\ell}{\epsilon}\right)^{-4}\left[K^{(0)}+\left(\frac{\epsilon}{\ell}\right)^{d-2}K^{(d-2)}\right]\,.
\end{align}
\end{subequations}
\endgroup
Moreover one finds $C_{\mu\nu\rho\sigma}C^{\mu\nu\rho\sigma}=K$, to leading order. The explicit expressions for the perturbative corrections to the curvature scalars are too lengthy to print them here. From equations~\eqref{4D brane horizon radius},~\eqref{4D brane horizon radius} and~\eqref{brane curvature scalars} one sees that $M^2R\propto {R_4}^2$ and $M^4R_{\mu\nu}R^{\mu\nu}\propto {R_4}^4$ in the 4-dimensional case, while $MR\propto {R_5}^3$ and $M^2R_{\mu\nu}R^{\mu\nu}\propto {R_5}^6$ in the 5-dimensional case.
\section{Numerical Results}\label{sec: numerical results}
To enable us to solve the ODEs we derived numerically the authors of~\cite{Figueras:2011va,Figueras:2014dta} kindly provided their data for the functions $s_4,\,t_4$ of equations~\eqref{4D CFT stress tensor} and~\eqref{5D CFT stress tensor}. In the 5-dimensional case data for the CFT stress-energy tensor are available for $\kappa\mu^{1/2}=n/16,\,n=1\ldots 16$. We set $G_d=\ell=1$ for our numerical calculations. As $\kappa>0$ for all data sets, the boundary metric is never extremal. All calculations were performed with the help of {\it Mathematica}, using both a finite differences algorithm and a pseudo-spectral algorithm. The latter showed better convergence (see appendix~\ref{app: numerical error} for details) and was hence used for all results presented in this paper. The available stationary CFT data only allows us to work on an unstructured grid in the $(J_5,R_5)$ parameter space, as explained in more detail in appendix~\ref{app: numerical error}. Interpolation on unstructured grids is challenging and {\it Mathematica} can only handle it to first order. In the plots all quantities have been rescaled to make them dimensionless. Where dotted lines have been added they are merely meant to guide the eye.
\subsection{Accuracy of the Leading Order Perturbative Construction}
\begin{figure}
\centering
\includegraphics[width=.85\textwidth]{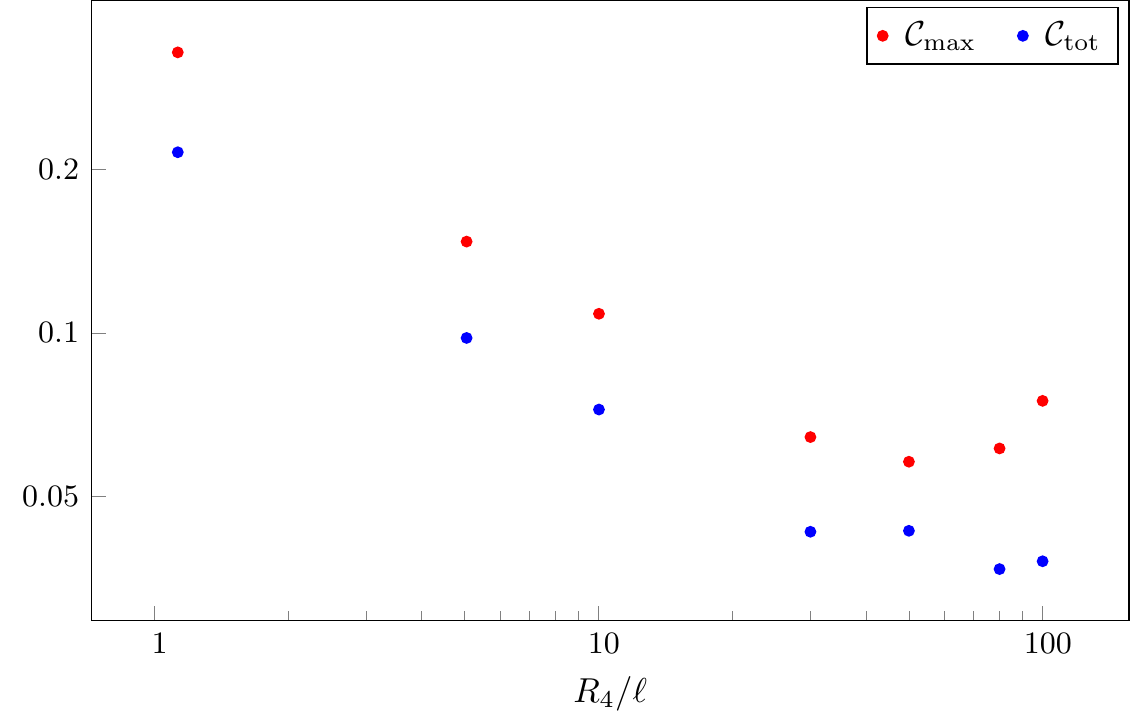}
\caption{The quantities $\mathcal{C}_1$, defined in~\eqref{compare max}, and $\mathcal{C}_2$, defined in~\eqref{compare tot}, indicate how well the leading order perturbative results approximate the full bulk solution depending on the horizon radius of the brane black hole.}
\label{fig:4Dcompare}
\end{figure}

By their perturbative nature our results are only reliable for small values of $\epsilon$, when the rescaled horizon radii $R_4/\ell$~\eqref{4D brane horizon radius} and $R_5/\ell$~\eqref{5D brane horizon radius} are large. The full static 5-dimensional bulk solution, whose brane metric our static results approximate, was constructed in~\cite{Figueras:2011gd,Figueras:2014nza} for horizon radii up to $100\ell$. By comparing it numerically to our results we can assess the range in which the leading order approximation can be trusted. We expect the results of the comparison to translate to the 5-dimensional case, as well as to higher-dimensional branes. As indicator of agreement we chose the difference in the square of the Weyl tensor, $\Delta C^2=C^2_{\text{full solution}}-C^2_{\text{leading order}}\,$, $C^2=C_{\mu\nu\rho\sigma}C^{\mu\nu\rho\sigma}$, while ${C^{(0)}}^2_H={C^{(0)}}^2(r=0)$, served as scale to determine when this difference is considered small. We calculated
\begin{equation}\label{compare max}
\mathcal{C}_{\rm max}\defeql\left(\frac{\max\left|\Delta C^2\right|}{\left|{C^{(0)}}^2_H\right|}\right)^{1/4}\,,
\end{equation}
and numerically integrated $\Delta C$ over the compact radial coordinate $r$ using {\it Mathematica}'s NIntegrate routine to obtain
\begin{equation}\label{compare tot}
\mathcal{C}_{\rm tot}\defeql\left(\frac{\int_{0}^{1}dr\left|\Delta C^2\right|}{\left|{C^{(0)}}^2_H\right|}\right)^{1/4}\,.
\end{equation}
Figure~\ref{fig:4Dcompare} shows both quantities as functions of the horizon radius. 
For large values of $R_4/\ell$ the Weyl tensor of the full solution is more seriously contaminated by the numerical errors and our comparison is  hence less accurate. As $\Delta C^2$ is largest at the horizon $\mathcal{C}_{\rm max}$ only samples the data points with the largest numerical error and is thus a less reliable indicator than $\mathcal{C}_{\rm tot}$.
\subsection{Metric and the Horizon Area}
\begin{figure}
\centering
\includegraphics[width=\textwidth]{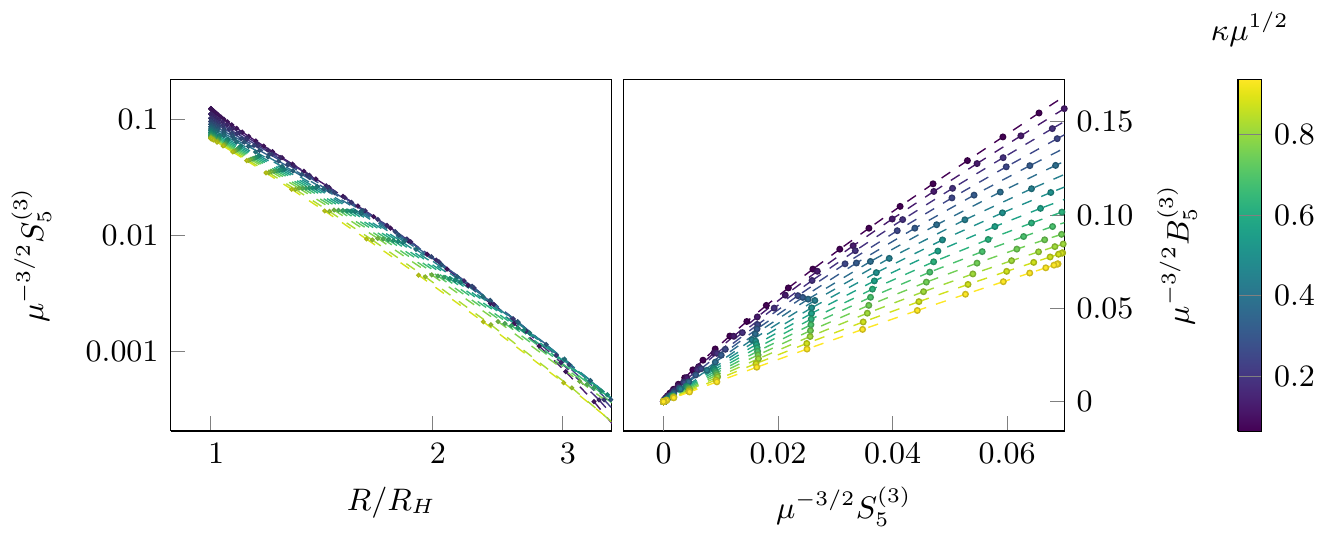}
\caption{This plot shows the change to the radius of the stationary metric (left), as well as the deformation of its spherical part as function of the change in radius (right).}
\label{fig:5Dmetricperturb}
\end{figure}

\begin{figure}
\centering
\includegraphics[width=\textwidth]{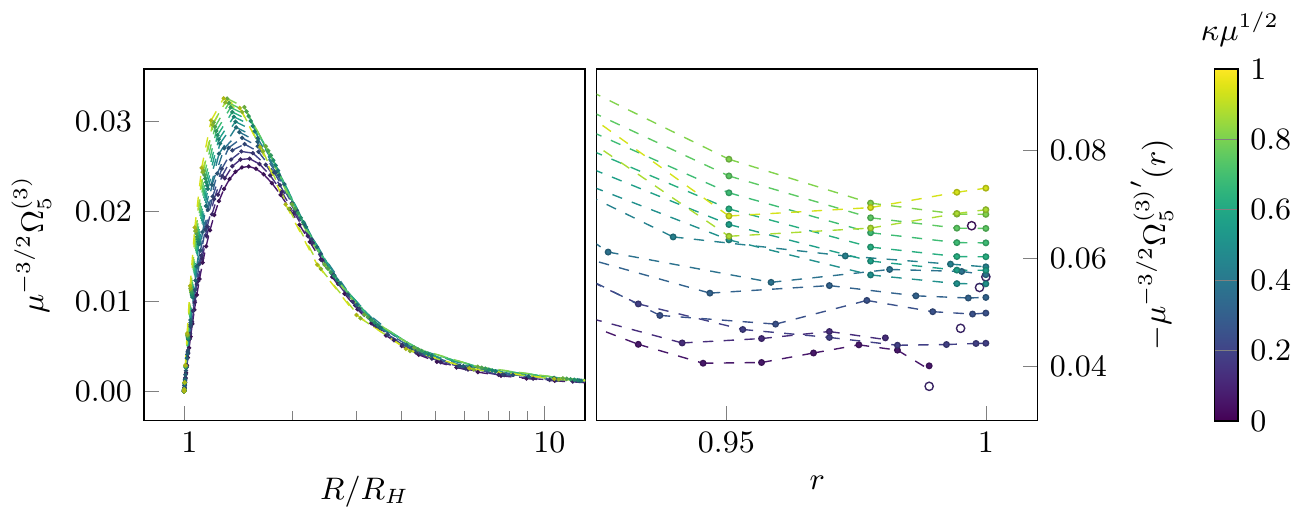}
\caption{The changes to both, the angular velocity (left) and the total angular momentum (right) are positive. Empty circles denote plot points discarded due to a lack of smoothness, as explained in appendix~\ref{app: numerical error}.}
\label{fig:5DD1Omega}
\end{figure}

\begin{figure}
\centering
\includegraphics[width=\textwidth]{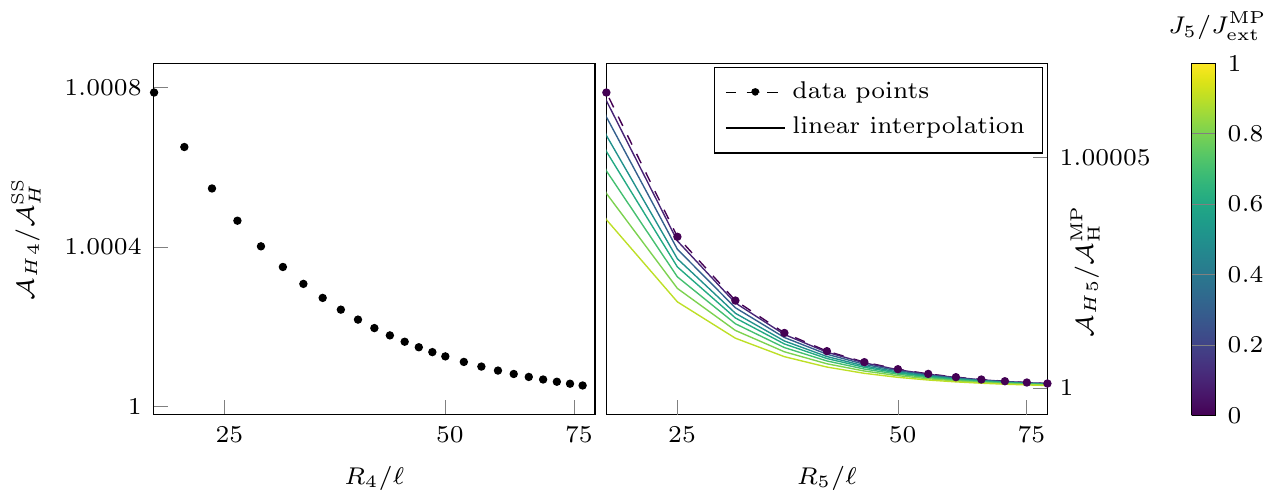}
\caption{Compared to a Schwarzschild or Myers-Perry solution the brane black holes have an enlarged horizon area. This effect is less pronounce for solutions with higher angular momentum.}
\label{fig:horizonarea}
\end{figure}

$S_4^{(2)}(r)$, which is the perturbative correction to the radius of the $SO(3)$ orbits of the static brane metric~\eqref{ansatz 4D}, is positive at $r=0$ and the mass~\eqref{brane BH mass} receives no perturbative correction. Hence the horizon area~${\mathcal{A}_4}_H$ (see equation~\eqref{4D horizon area}) of the brane black hole is, at equal mass, larger than the area $\mathcal{A}_{\rm H}^{\rm SS}$ of a Schwarzschild black hole (see left plot of figure~\ref{fig:horizonarea}). Figure~\ref{fig:5Dmetricperturb} shows the functions $S_5^{(3)}(r)$ and $B_5^{(3)}(S_5^{(3)})$, which determine the radius and the deformation of the angular part of the stationary brane metric~\eqref{ansatz 5D}. 
As the angular momentum $J_5$ itself receives a correction, one cannot immediately conclude how the horizon area ${\mathcal{A}_5}_H$ (see equation~\eqref{5D horizon area}) compares to the horizon area $\mathcal{A}_{\rm H}^{\rm MP}$ of a Myers-Perry black hole of the same mass and angular momentum. The function $\Omega_5^{(3)}(r)$ determines the correction to the angular velocity~\eqref{5D perturbations Omega} and its first derivative at infinity the change in the angular momentum~\eqref{5D brane angular momentum}. Both quantities are shown in figure~\ref{fig:5DD1Omega}.
Clearly, the correction to the brane angular momentum becomes smaller as the rotation of the background metric grows. If the correction remains positive for background metrics very close to extremality, large brane black holes that violate the Myers-Perry extremality bound could be constructed.
The left plot of figure~\ref{fig:horizonarea} shows how the horizon area of the stationary brane black holes compares to a Myers-Perry black hole.
\subsection{Curvature Scalars}
\begin{figure}
\centering
\includegraphics[width=\textwidth]{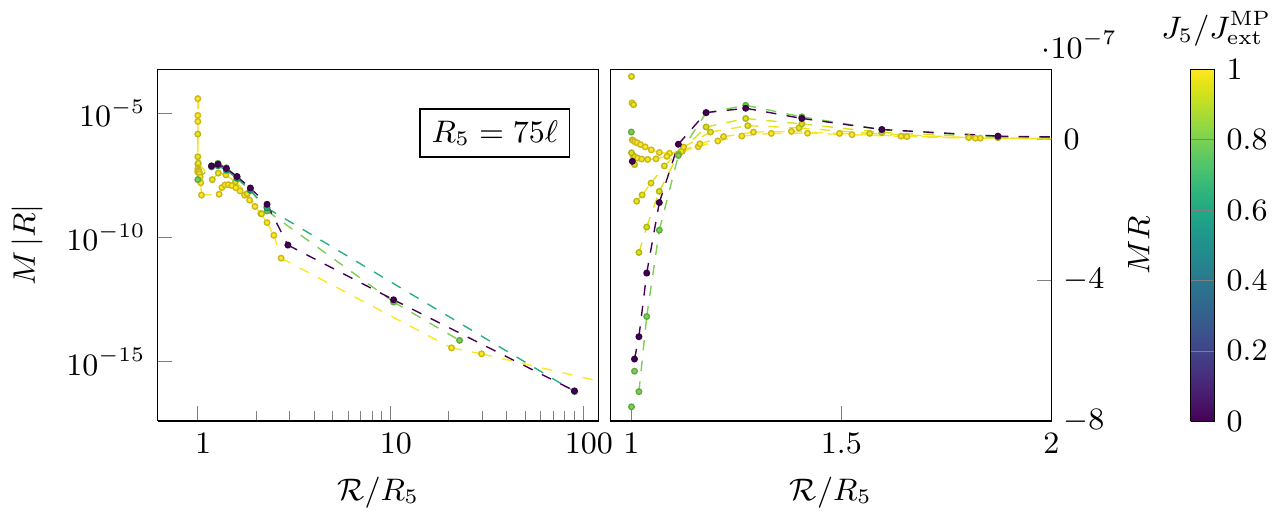}
\caption{The Ricci scalar of our 5-dimensional black hole solutions at fixed horizon radius.}
\label{fig:5Dscalar}
\end{figure}

\begin{figure}
\centering
\includegraphics[width=\textwidth]{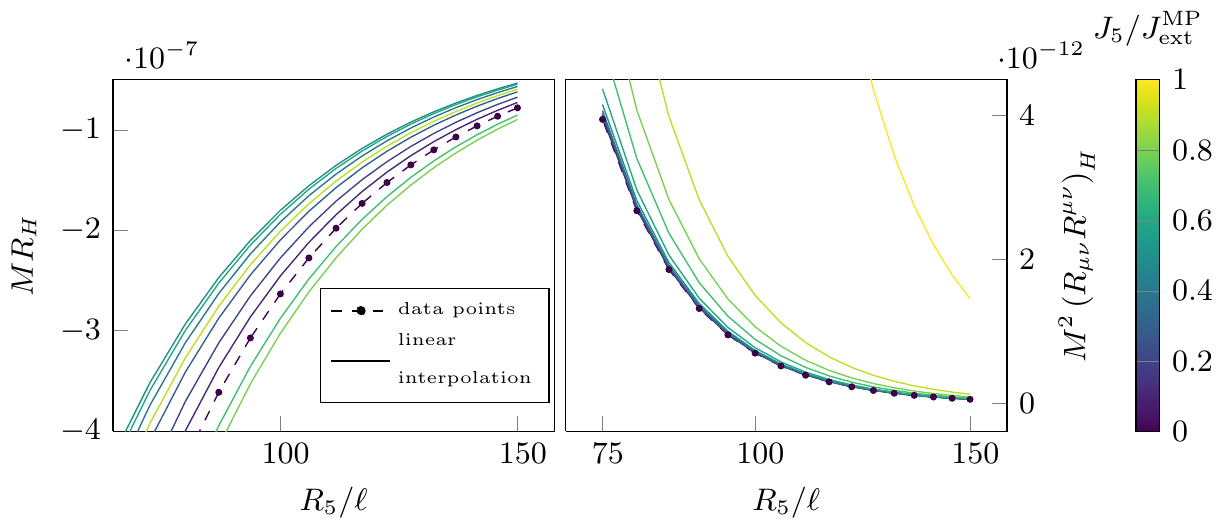}
\caption{The horizon value of the Ricci scalar and the square of the Ricci tensor of our 5-dimensional black hole solutions as function of the horizon radius.}
\label{fig:5DscalarHricciH}
\end{figure}

\begin{figure}
\centering
\includegraphics[width=\textwidth]{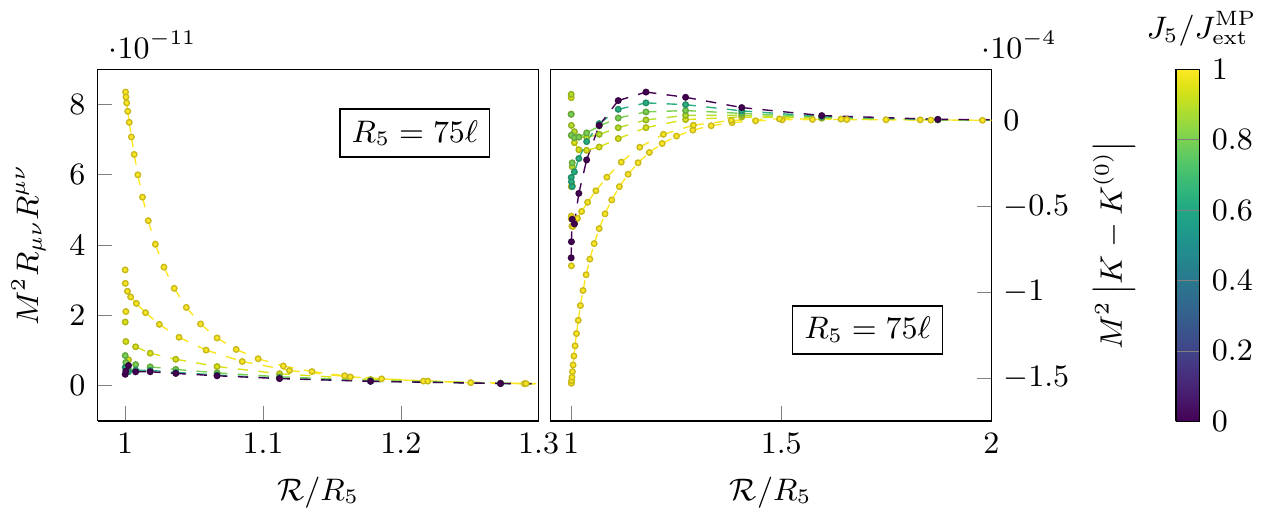}
\caption{The square of the Ricci scalar and the corrections to the Kretschmann scalar for our 5-dimensional solutions as function of the horizon radius.}
\label{fig:5Driccisquarekretschmann}
\end{figure}

\begin{figure}
\centering
\includegraphics[width=\textwidth]{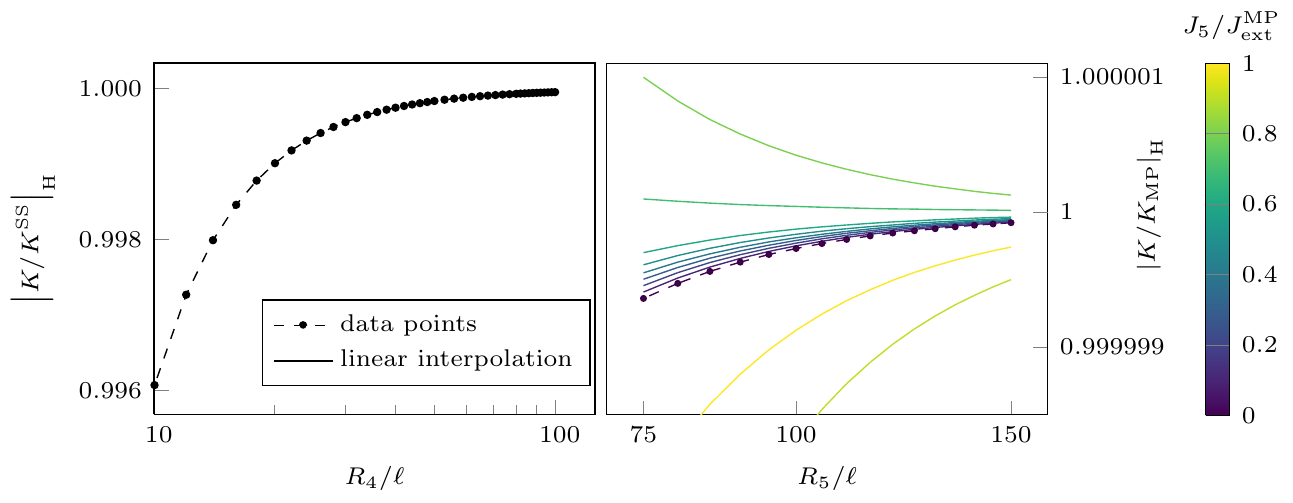}
\caption{The Kretschmann scalar of our black hole solutions in 4 and 5 dimensions compared to a Schwarzschild or Myers-Perry black hole of the same mass and angular momentum.}
\label{fig:riemannratio}
\end{figure}

The static curvature scalars of our solutions show the same behaviour in 4 and 5 dimensions. So, to avoid redundancy, most plots in this section only show the 5-dimensional results. We plotted the curvature scalars of equation~\eqref{brane curvature scalars} as functions of a non-compact radial coordinate $\mathcal{R}$, defined as
\begin{equation}\label{4D my non-compact coordinate}
\mathcal{R}\defeql\frac{R_4}{1-r^2}\,,
\end{equation}
in the 4-dimensional case, and
\begin{equation}\label{5D my non-compact coordinate}
\mathcal{R}^2+\left(\frac{3 J_5}{2 M_5}\right)^2\defeql\frac{R_5^2+\left(\frac{3 J_5}{2 M_5}\right)^2}{\left(1-r^2\right)^2}
\end{equation}
in the 5-dimensional case. Note that $a=\frac{3 J}{2 M}$ for a five-dimensional equal angular momenta Myers-Perry black hole. So, in the limit $\epsilon\rightarrow 0$ definitions~\eqref{4D my non-compact coordinate} and~\eqref{5D my non-compact coordinate} reduce to the compact radial coordinates~\eqref{4D compact coordinate} and~\eqref{5D compact coordinate}. For the plots we chose $R_{4/5}=75\ell$. Another value of $R_{4/5}$ would not change the results qualitatively.  We also plotted the value of the curvature scalars at the horizon as the black hole parameters vary. The corrections to the background scalars are small enough to be significantly spoiled by numerical errors near the horizon up to $\mathcal{R}/R_{4/5}\sim 1.01$, as illustrated by the left plot of figure~\ref{fig:5Dscalar}. The error is most serious so for the fast spinning data sets. To obtain realistic results for the values of the curvature scalars very close to the horizon, we had to disregard the first few grid points near $r=0$. The curvature scalars are largest at the horizon and approach zero rapidly as the radial coordinate grows, as shown exemplary for the 5-dimensional Ricci scalar on the left of figure~\ref{fig:5Dscalar}. The Ricci scalar is negative at the horizon for lower values of the angular momentum and becomes positive for high rotation (see right plot of figure~\ref{fig:5Dscalar}), whereas the square of the Ricci tensor is always positive at the horizon (see left plot of figure~\ref{fig:5Driccisquarekretschmann}). Figure~\ref{fig:5DscalarHricciH} illustrates that $R\propto {R_4}^3$, $R_{\mu\nu}R^{\mu\nu}\propto {R_4}^6$ in the 5-dimensional case.\\
As both $R$ and $R_{\mu\nu}$ vanish on a Schwarzschild or Myers-Perry background, the Kretschmann scalar is the only curvature scalar that allows a direct comparison between the brane black holes and black holes of the same mass and angular momentum in pure general relativity. (Recall that to leading order $C_{\mu\nu\rho\sigma}C^{\mu\nu\rho\sigma}=K$ for our solutions.) The perturbative corrections to the 5-dimensional background Kretschmann scalar are shown in the left plot of figure~\ref{fig:5Driccisquarekretschmann}. The corrections are small compared to $K^{(0)}$, so the behaviour of the full Kretschmann scalar as function of the radial coordinate is dominated by $K^{(0)}$. Lastly, figure~\ref{fig:riemannratio} shows that, for vanishing, small, and very large angular momenta, the value of the Kretschmann scalar at the horizon is smaller than that of a Schwarzschild or Myers-Perry black hole, whereas for medium angular momentum it is larger.
\section*{Acknowledgments}
Special thanks go to Pau Figueras, whose advice was essential to the research for this publication. It is a pleasure to thank Malcolm J. Perry and Saran Tunyasuvunakool for helpful discussions. MS acknowledges financial support from the British Engineering and Physical Sciences Research Council (EPSRC), the German Academic Exchange Service, and the Cambridge European Trust. 

\begin{appendix}
\section{Gravitational Equations on the Brane}\label{app: grav eqs brane}
This section closely follows~\cite{deHaro:2000wj,Figueras:2011gd}, where all formulae and details of the derivation can be found. Note, however, that we use a different sign convention for the Riemann tensor than~\cite{deHaro:2000wj}.\\
Near the conformal boundary any $(d+1)$-dimensional asymptotically AdS metric can be brought into the form~\cite{Fefferman:1985gws,Fefferman:2007rka} 
\begin{equation}\label{FG expansion}
\begin{aligned}
&ds^2=g_{AB}dx^A dx^B=\frac{\ell^2}{z^2}\Bigl(dz^2+\tilde{g}_{\mu\nu}(z,x)dx^\mu dx^\nu\Bigr)\,,\\
&\tilde{g}_{\mu\nu}(z,x)=g^{(0)}_{\mu\nu}(x)+z^2g^{(2)}_{\mu\nu}(x)+\ldots+z^dg^{(d)}_{\mu\nu}(x)+z^d\log z^2 h^{(d)}_{\mu\nu}(x)+\mathcal{O}\left(z^{d+1}\right)\,,
\end{aligned}
\end{equation}
where $g^{(0)}_{\mu\nu}$ is a representative of the boundary conformal structure. Up to order $(d-1)$ only even powers of $z$ appear in the expansion and the logarithmic term is absent in odd dimensions. The Einstein equations uniquely determine the coefficients $g^{(2k)}_{\mu\nu}$, $k\in\Naturals,k<d/2$, and $h^{(d)}_{\mu\nu}$ as functions of $g^{(0)}_{\mu\nu}$. The series~\eqref{FG expansion} is a curvature expansion, where the coefficients $a^{(2k)}$ are of order $R\left[g^{(0)}\right]^k$. To determine the remaining coefficients, a symmetric, covariantly conserved tensor $t_{\mu\nu}$, which arises as an integration constant at order $d$, must be specified. In general 
\begin{equation}\label{integration constant t}
g^{(d)}_{\mu\nu}=t_{\mu\nu}+\chi^{(d)}_{\mu\nu}\left[g^{(0)}\right]\,,\qquad \chi^{(2k+1)}_{\mu\nu}\left[g^{(0)}\right]=0\,
\end{equation}
and $t_{\mu\nu}$ is related to stress-energy tensor of the dual CFT via
\begin{equation}\label{general form of stress tensor}
\langle T^{\text{CFT}}_{\mu\nu}\rangle=\frac{d \ell^{d-1}}{16\pi G_{d+1}}g^{(d)}_{\mu\nu}+X^{(d)}_{\mu\nu}\left[g^{(0)}\right]\,,\qquad X^{(2k+1)}_{\mu\nu}\left[g^{(0)}\right]=0\,.
\end{equation}
The bulk gravitational equations imply that tensor is covariantly conserved with respect to $g^{(0)}_{\mu\nu}$ and its trace reproduces the conformal anomaly of the dual CFT.\\
Now consider a brane placed close to the conformal boundary at $z=\epsilon$, where the expansion~\eqref{FG expansion} is valid. Hypersurfaces of constant $z$ have normal vector $n^A=-\frac{z}{\ell}\delta^A_z$, extrinsic curvature
\begin{equation}\label{extrinsic curvature brane}
K_{\mu\nu}=\frac{1}{2}\,\mathcal{L}_n \gamma_{AB}=-\frac{z}{2\ell}\partial_z\left(\frac{\ell^2}{z^2}\,\tilde{g}_{\mu\nu}\right),
\end{equation}
and curvature radius
\begin{equation}
R_c\propto \frac{1}{\sqrt{\left|K_{\mu\nu}\right|}}=\frac{z}{\sqrt{\ell\,\left|g^{(0)}_{\mu\nu}\right|}}+\mathcal{O}\left(z^5\right).
\end{equation}
Hence the parameter $\epsilon$ controls the curvature radius of the brane compared to the bulk AdS length. The gravitational dynamics on the brane are governed by the Israel equations
\begin{equation}\label{Israel eqs brane}
K_{\mu\nu}(z=\epsilon)-K(z=\epsilon)\gamma_{\mu\nu}=4\pi G_{d+1}\left(-\sigma\gamma_{\mu\nu}+\tau_{\mu\nu}\right)\,,
\end{equation}
where $\sigma$ is the brane tension and $\tau_{\mu\nu}$ the stress-energy tensor of matter on the brane. To derive the gravitational equations on the brane substitute expression~\eqref{extrinsic curvature brane} for the extrinsic curvature in~\eqref{Israel eqs brane}. Then perturbatively invert the relationship between $g^{(0)}_{\mu\nu}$ and $\tilde{g}_{\mu\nu}$ to produce\footnote{The notation $T_{\mu\nu}[\gamma]$ is somewhat lax. It labels the stress-energy tensor of the theory on the hypersurface $g_{\mu\nu}=\gamma_{\mu\nu}$, but does not imply that the tensor is given as functional of $\gamma$.}
\begin{equation}\label{RS2 brane grav eq_AdS/CFT}
G_{\mu\nu}[\gamma]+\lambda\gamma_{\mu\nu}+\mathcal{O}\left(R[\gamma]^2\right)=8\pi\,G_d\Bigr[\tau_{\mu\nu}+2\langle T^{\text{CFT}}_{\mu\nu}[\gamma]\rangle\Bigr]\,,
\end{equation}
with
\begin{subequations}\label{Newton & cosmo const on brane AdS/CFT}
\begin{align}
G_d=&\,\frac{\left(d-2\right)G_{d+1}}{2\ell}\,\label{Newton const brane AdS/CFT},\\
\lambda=&-\,\frac{\left(d-1\right)\left(d-2\right)}{\ell^2}+8\pi G_d\,\sigma\,.\label{cosmo const brane AdS/CFT}
\end{align}
\end{subequations}
 The coefficients up to order $d$ in~\ref{FG expansion} plus the logarithmic term give rise to higher order curvature corrections indicated by $\mathcal{O}\left(R[\gamma]^2\right)$, while the coefficient $g_{\mu\nu}^{(d)}$ provides $\langle T^{\text{CFT}}_{\mu\nu}[\gamma]\rangle$ (as well as additional curvature corrections). Contributions from coefficients of order higher than $d$ involve derivatives of the stress-energy tensor.
\section{Numerical Errors and Consistency Checks}\label{app: numerical error}
\subsection{Details of the Numerical Caluclations}
\begin{figure}
\centering
\includegraphics[width=.85\textwidth]{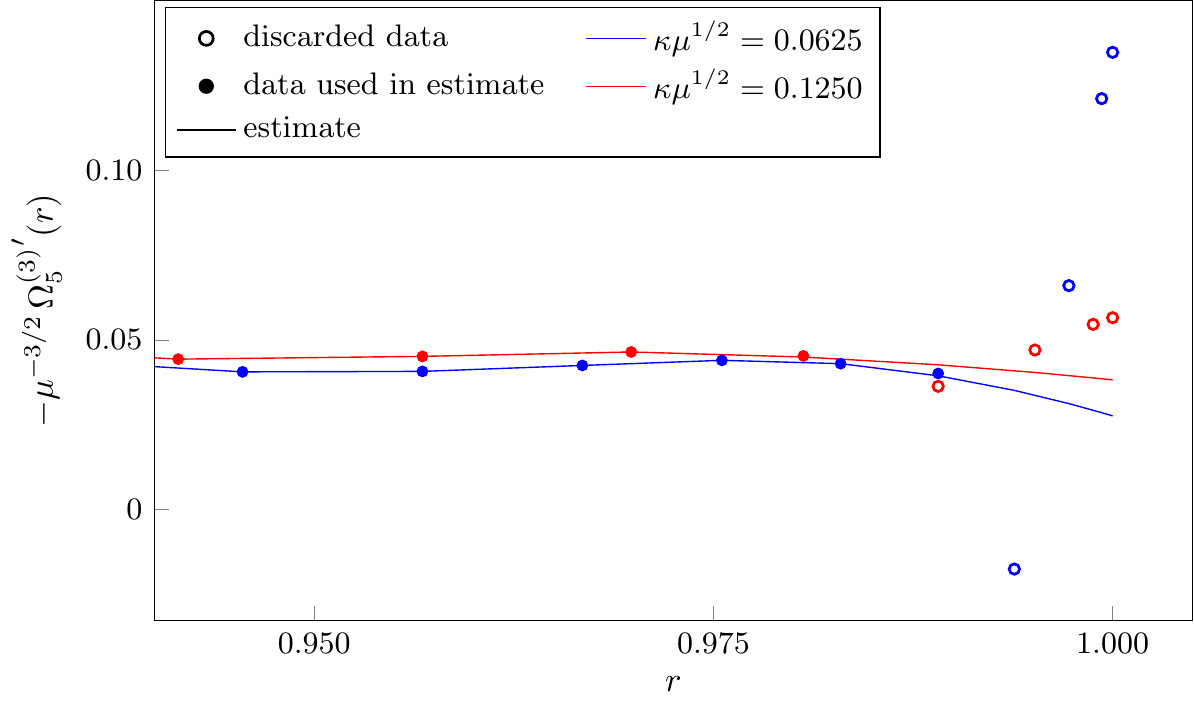}
\caption{Due to a lack of smoothness in the CFT data our results for ${\Omega_5^{(3)}}'$ near $r=1$ for the two fastest spinning data sets are not reasonable. As illustrated in this plot we chose to estimate the value of ${\Omega_5^{(3)}}'(1)$ via interpolation.}
\label{fig:5DD1Omegacorrected}
\end{figure}

\begin{figure}
\centering
\includegraphics[width=\textwidth]{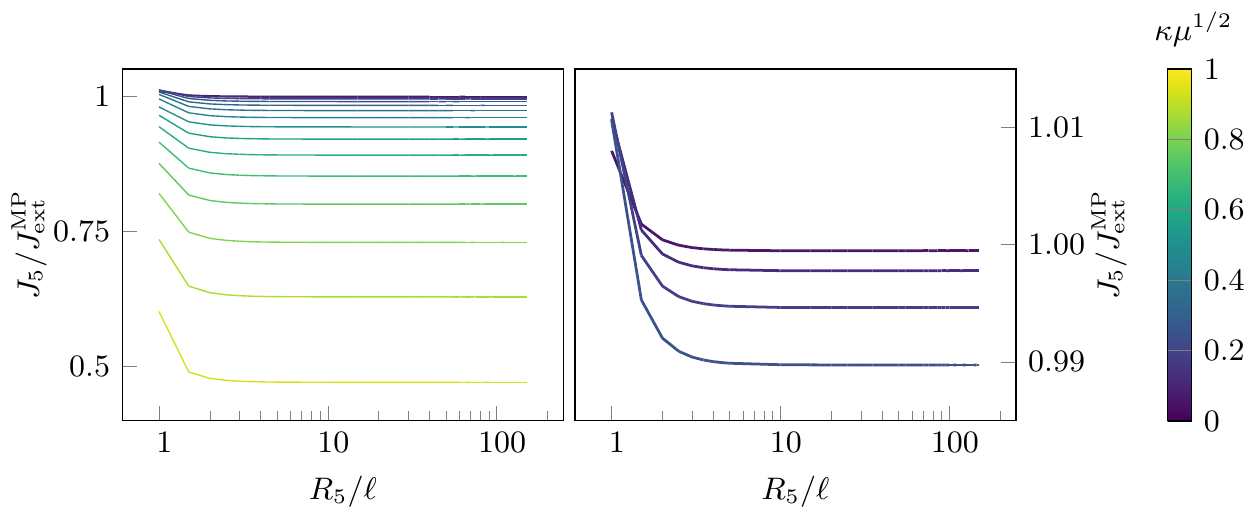}
\caption{This plot of $J_5$ as a function of $R_5$ shows which subset of the parameter space the available CFT data enables us to access.}
\label{fig:5DDJofR5}
\end{figure}

 After imposing the boundary conditions detailed in the section~\ref{subsec: gauge & bcs} and discretizing the ODEs on a Chebyshev grid the resulting matrix equation was solved with {\it Mathematica}'s LinearSolve routine. In the 5-dimensional case, however, this did not yield sufficiently smooth data for ${\Omega_5^{(3)}}'$ near $r=0$. This problem occurs independently of the algorithm we use and stems from a lack of smoothness in the CFT data, not from an error in our calculation. Imposing an additional boundary condition,
\begin{equation}
{\Omega_5^{(3)}}''(r)|_{r=1}=0\,,
\end{equation}
significantly improved the results, but did not entirely remedy the issue: The results for the two fastest-spinning stationary CFT data sets still showed a serious lack of smoothness, while the rest of the data was almost smooth. According to equation~\eqref{5D brane angular momentum} the value of ${\Omega_5^{(3)}}'(1)$ determines the correction to the angular momentum $J_5$. Though, for large horizon radii this correction is very small compared to the angular momentum $J_0$ of the background metric. So small inaccuracies in ${\Omega_5^{(3)}}'$ will not significantly change the nature of our results. Hence we chose to estimate the correct value of ${\Omega_5^{(3)}}'(1)$ for the two fastest-spinning stationary CFT data sets only. To do so we discarded the last few grid points and interpolated the remaining data with {\it Mathematica}'s Integrate routine, as illustrated in figure~\ref{fig:5DD1Omegacorrected}.\\
As mentioned in section~\ref{sec: numerical results} the available CFT data only allows us to determine our stationary results on an unstructured grid in the $(R_5,J_5)$ parameter space. The reason for this is the following: A pair of values for $\epsilon$ and $\kappa\mu^{1/2}$ of the background metric corresponds to a point in the $(R_5,J_5)$ parameter space. As CFT data is only available for $\kappa\mu^{1/2}=n/16,\,n=1\ldots 16$, we can merely calculate results for a subset of the parameter space. Given a CFT data set with fixed $\kappa\mu^{1/2}$, as $\epsilon$ varies one travels along a curve through the $(R_5,J_5)$ parameter space, as illustrated in figure~\ref{fig:5DDJofR5}.
 To probe the space of all stationary solutions, we would ideally like to work on a regular $(R_5,J_5)$-grid, on which one can use standard interpolation routines. For large horizon radii, however, any grid completely contained in the accessible subset of the parameter space is irregular.
\subsection{Numerical Errors}
\begin{figure}
\includegraphics[width=.85\textwidth]{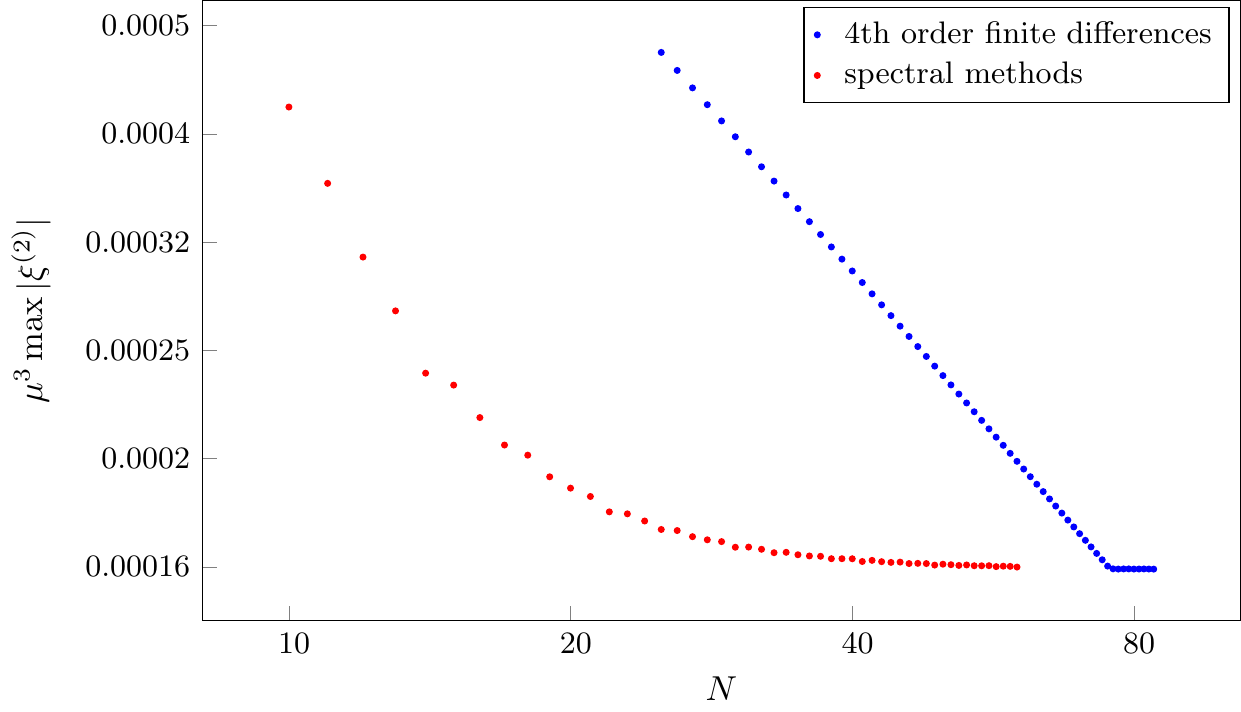}
\caption{A doubly logarithmic plot of the maximum norm of the DeTurck vector field as function of the grid size confirms that the overall numerical error stagnates for large enough $N$ at a minimum value, which is determined by the error of the CFT data and thus independent of the algorithm.}.
\label{fig:4Dxi}
\end{figure}

\begin{figure}
\begin{flushleft}
\includegraphics[width=\textwidth]{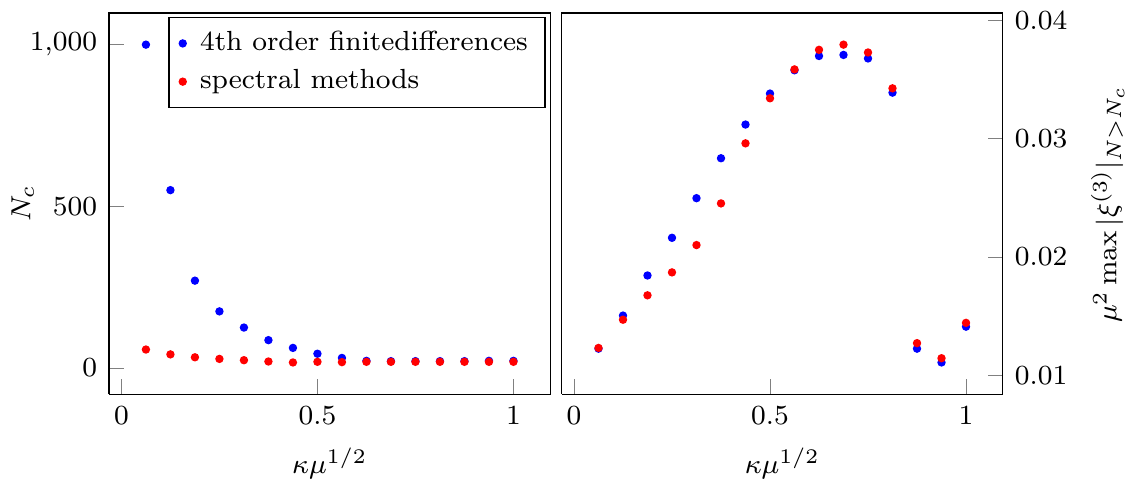}
\caption{The threshold value $N_c$ beyond which the overall numerical error stagnates increases with the rotation of the background metric, more significantly so for the finite differences algorithm. The numerical error varies with the rotation of the background metric, but also depends on the grid used to determine the CFT data. (A finer grid was used  for the three slowest spinning cases, as well as for the 4-dimensional case.)}
\label{fig:5Dconvergence}
\end{flushleft}
\end{figure}

\begin{figure}
\centering
\includegraphics[width=.85\textwidth]{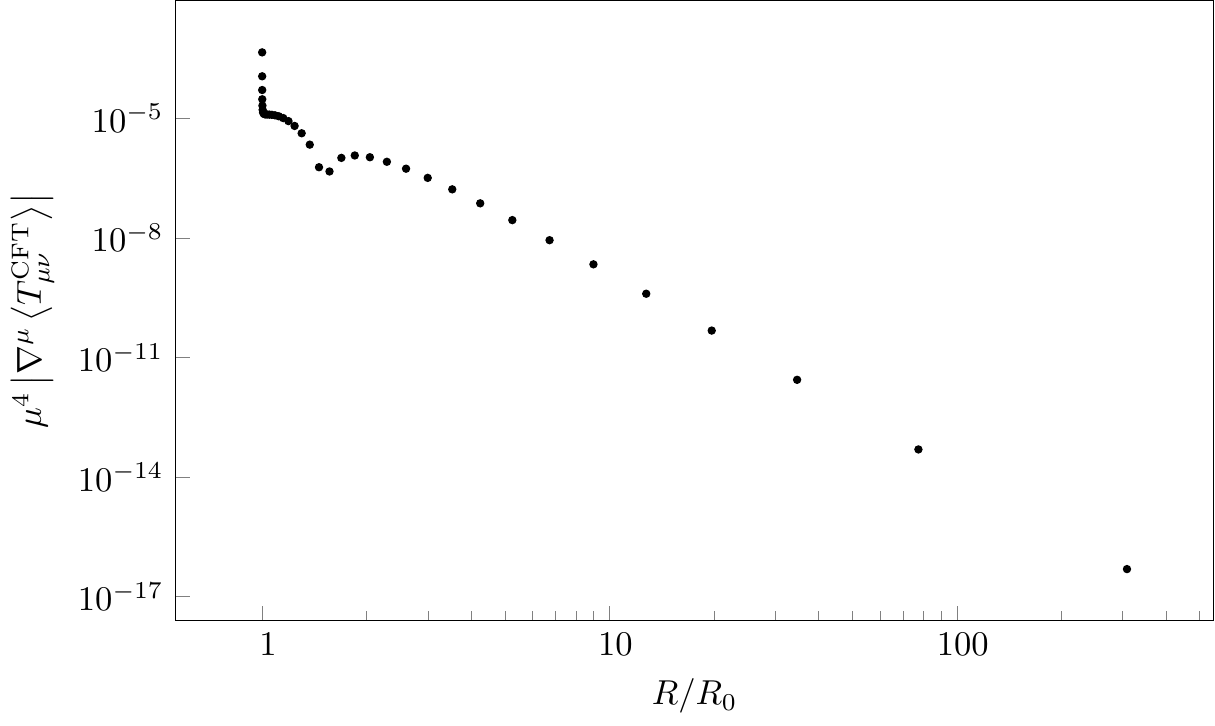}
\caption{The divergence of the static CFT stress tensor as function of the radial coordinate shows that the numerical error of the CFT data is largest at $R=R_0$ and decreases rapidly as one moves away from the horizon.}
\label{fig:4Dstressdiv}
\end{figure}

Two factors contribute to the overall numerical error of our results: the discretization error of the grid and the error of the CFT data we received. The former decreases as the grid size $N$ grows while the latter remains fixed. The DeTurck gauge condition $\xi^\mu=0$ can serve as a measure for the overall error because it is solved numerically along with the main equations. 
	In~\cite{Figueras:2011va,Figueras:2013jja} the CFT data was determined without imposing the constraints, which encode tracelessness and covariant conservation of the CFT stress tensor. Consequently the numerical value of the constraints provides a measure for the error of the CFT data. For small $N$ the overall numerical error will be dominated by the discretization error, i.e., it decreases as the grid size grows. Beyond a threshold value $N_c$ the error of the CFT data will dominate causing the overall error to stagnate. To achieve the best accuracy possible for our results we need to work at $N>N_c$. The DeTurck vector field is of the form $\xi^\mu=\epsilon^{d-2}{\xi^{(d-2)}}^\mu$ and figure~\ref{fig:4Dxi} shows its norm as function of the grid size in the 4-dimensional case.
The numerical error in the 5-dimensional case behaves as expected, so figure~\ref{fig:5Dconvergence} merely shows $N_c$ and $\mu^2\max|\xi^{(3)}|_{N>N_c}$ for the different CFT data sets.	
The figures above clearly show, that the pseudo-spectral algorithm has better convergence properties than finite differences algorithms.\\ 
According to Figure~\ref{fig:4Dstressdiv}, a doubly logarithmic plot of the divergence of the static CFT stress tensor, the numerical error of the CFT data is largest at the horizon and decreases rapidly as the radial coordinate grows.
The same can be observed in the 5-dimensional case for both $\mu^{5/2}\left|\left\langle {T^{\rm CFT}}_\mu^\mu\right\rangle\right|$ and $\mu^{5/2}\left|\nabla^\mu\left\langle T^{\rm CFT}_{\mu\nu}\right\rangle\right|$. The overall numerical error of our results inherits this behaviour from the dominant  error of the CFT data. Consequently, small quantities that are of the same order of magnitude as the numerical error are seriously affected by noise near the horizon but still reliable for larger values of the radial coordinate.
\subsection{Consistency Checks}
\begin{figure}
\centering
\includegraphics[width=\textwidth]{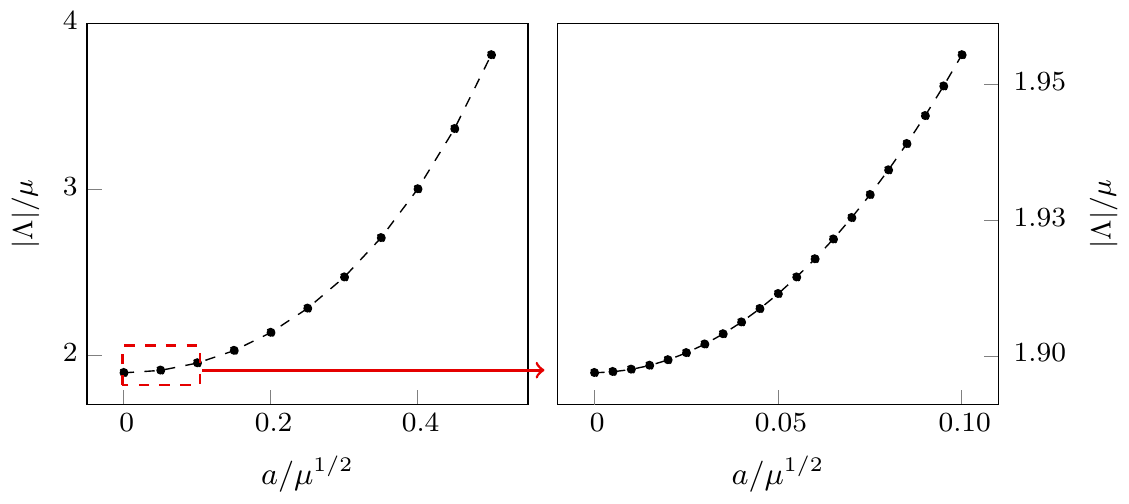}
\caption{Our calculations correctly show that the negative mode of the equal angular momenta Myers-Perry metric varies smoothly with the rotation parameter, hence affirming our code is free of errors. }
\label{fig:5DEVs}
\end{figure}

The linearized DeTurck gauge can be used to calculate the eigenvalues of the Lichnerowicz operator, by solving the equation
\begin{equation}
{R_{\mu\nu}^{\rm H}}^{(1)}=\Lambda h_{\mu\nu}\,.
\end{equation}
To check both the ODEs we obtained and our numerical calculations, we slightly modified our code to determine the negative modes of the background metric. We correctly reproduced $\Lambda=-0.19196/\mu^2$ for the Euclidean Schwarzschild metric~\cite{Gross:1982cv} and found that the negative eigenvalue of the 5-dimensional equal angular momenta Myers-Perry metric varies smoothly with the value of the angular momentum parameter $a$, as shown in figure~\ref{fig:5DEVs}.
\end{appendix}
\bibliographystyle{plain}
\bibliography{refs}

\end{document}